# Extension of MIH to Support FPMIPv6 for Optimized Heterogeneous Handover

*Jianfeng Guan[1], Vishal Sharma[2], Ilsun You[2*], Mohammad Atiquzzaman[3]*
*[1]State Key Laboratory of Networking and Switching Technology*
*Beijing University of Posts and Telecommunications, Beijing, China, 100876*
*jfguan@bupt.edu.cn*
*[2]Department of Information Security Engineering*
*Soonchunhyang University, Asan-si, Republic of Korea*
*isyou@sch.ac.kr,* vishal_sharma2012@hotmail.com
*[3]School of Computer Science, University of Oklahoma, Norman, OK,*
*Email: atiq@ou.edu*
**Corresponding author*

## Abstract

Fast handover for Proxy Mobile IPv6 (FPMIPv6) can reduce handover delay and packet loss compared with Proxy Mobile IPv6 (PMIPv6). However, FPMIPv6 still cannot handle heterogeneous handovers due to the lack of unified Layer 2 triggering mechanism along with the booming of emerging wireless technologies. Media Independent Handover (MIH) can provide heterogeneous handover support, and a lot of integration solutions have been proposed for it. However, most of them focus on the integration of MIH and PMIPv6, and require the additional mechanisms, which are out of the scope the MIH and difficult to standardize the operations. Therefore, in this paper, we propose an integration solution of FPMIPv6 and MIH by extending the existing MIH standards, and adopt the city section mobility model to analyze its performance under different scenarios. The analytical results show that the proposed solution is capable of reducing the handover delay and the signaling cost compared with the standard as well as the fast handover solutions.

## 1 Introduction

The massive development of wireless technologies such as Wi-FI, WiMAX, LTE, 5G promotes the booming of various mobile devices with multiple wireless network interfaces. The coexistence and mutual complementation of multiple wireless networks derive the demand for higher bandwidth with enhanced QoE and QoS. This has spurred the related researches of mobility management to support seamless mobility services for media rich applications.

According to Cisco report [1], the global mobile data traffic has grown 74% in 2015, and the expected mobile data traffic will be 30.6 EB/Month by 2020, in which mobile video content will occupy about 23 EB/Month. Mobile video has higher QoE requirements than data services, which makes the users tend to adopt high-speed wireless network with guaranteed QoS. Considering that more and more mobile devices have equipped multiple network interfaces, it is important for mobile users to select the most appropriate interface to improve the bandwidth and reduce the cost. However, due to the limited coverage, heterogeneous handover will happen when users leave from one wireless network to another. Keeping the continuous ongoing session during the handover is an urgent problem.

Various solutions have been proposed [3] to provide mobility for the emerging mobile devices, in which IP mobility management solutions have got more attentions and many schemes have been proposed [4]. These solutions can be classified into the host-based mobility and the network-based mobility in terms of handover initiator; single-host mobility and sub-network mobility according to the mobility types; and the central mobility management solutions and distributed mobility management in terms of the distribution of mobility-related entities. Among these solutions, Mobile IPv6 (MIPv6) [5] is most cited scheme published by IETF, in which Home Agent (HA) manages the registered Mobile Nodes (MNs) and maintains the bi-directional tunnel for each MN that leaves the home network. Due to the triangle routing problem, Hierarchical Mobile IPv6 (HMIPv6) [6] was proposed to divide topology into different domains to reduce the signaling cost for the MNs with micro-mobility. In HMIPv6, Mobility Anchor Point (MAP) is introduced as the "HA" of the given domain to reduce MNs' updating cost. Another important improvement is the Mobile IPv6 Fast Handovers (FMIPv6) [7], which reduce the handover delay and packet loss with the aid of link layer triggering and pre-registration mechanisms. More specially, FMIPv6 sets up a tunnel between previous attachment point and new attachment point to forward buffered packets during handover to reduce packet loss. However, these solutions require the involvement of MNs which may result in excessive energy consumption for energy-limited MNs. Therefore, Proxy MIPv6 (PMIPv6) [13] is proposed by introducing Mobile Access Gateway (MAG) to perform the mobility-related operations on behalf of MNs, and it can reduce handover delay and signaling cost [14] [15] compared with the previous solutions. PMIPv6 can be further improved by combining the fast handover mechanism which is also known as the Fast handover for PMIPv6 (FPMIPv6) [16]. However, the basic FPMIPv6 does not provide the support for heterogeneous handovers.

For MN with dual radios, Liehbsch *et al.* [8] introduce a new transient binding mobility option in PBU/PBA, which sets up a temporary Binding Cache Entry (BCE) to forward packets to PMAG and NMAG simultaneously, which can avoid superfluous packet forwarding delay or even packet loss. Transient binding can be initiated by NMAG or LMA, and some studies [9][10] have adopted it to improve handover performance. Due to the adoption of multiple interfaces, it is easy to support heterogeneous handover.

Besides, IEEE has published Media Independent Handover (MIH) framework (IEEE-802.21) to support heterogeneous handover. MIH can optimize the network selection, realize the seamless roam, and lower the power consumption. The previous heterogeneous handover schemes [17] generally combine the PMIPv6 and MIH, while in this paper, we propose an integration model of FPMIPv6 and MIH to further improve handover performance. The main contributions are summarized as follows.

(1) An integration model of FMIPv6 and MIH is proposed.

(2) MIH TLV is extended to contain the related information used for pre-registration without introducing any additional signaling messages.

The rest of this paper is organized as follows. Section 2 introduces FPMIPv6, MIH, and investigates the related work. Section 3 describes the proposed integration model of FPMIPv6 and MIH. Section 4 evaluates its performance and compares it with the related solutions. Section 5 summarizes this work and analyzes the potential challenges and further research directions.

# 2 Related works

## 2.1 Overview of FPMIPv6

FPMIPv6, as an enhancement of PMIPv6, adopts pre-registration and short time tunnel to reduce the handover delay and the packets loss. Figure 1 shows atypical reference network model of FPMIPv6, which consists of LMA, PMAG, NMAG, P-AN and N-AN.LMA is the topological anchor point that performs home agent function for each registered MNs and manages MN's binding and reachability.

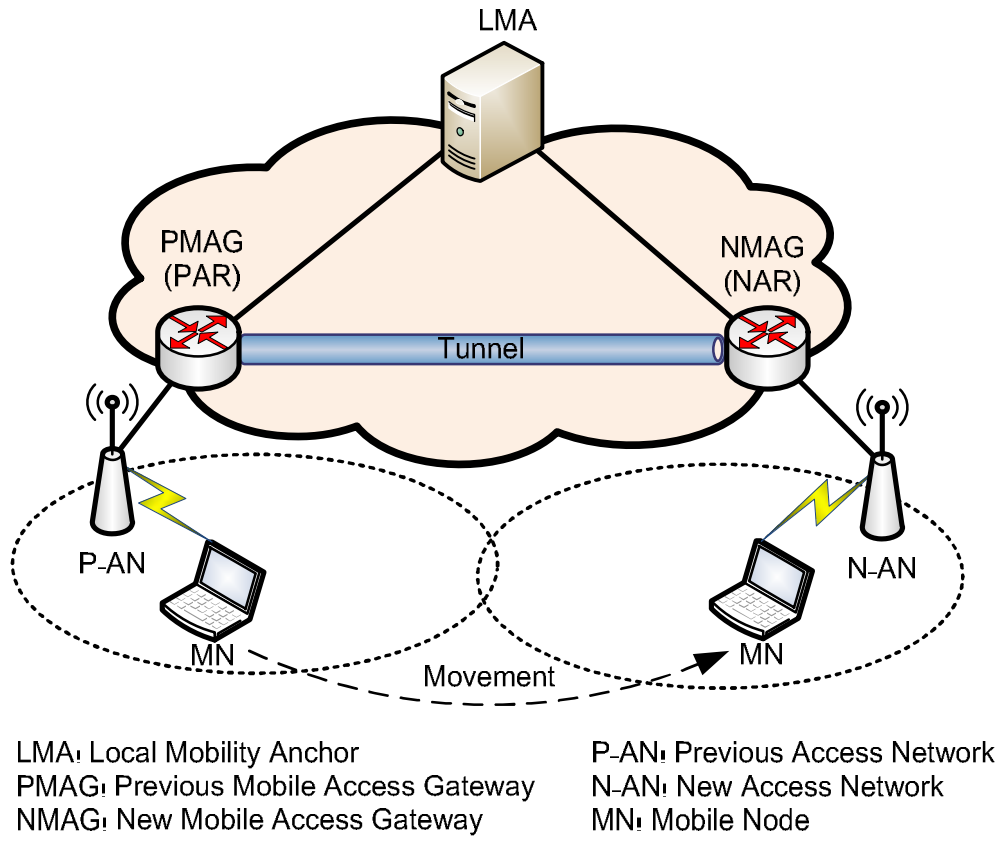

Figure 1 The reference network model of FPMIPv6

MAG is a function on an access router to manage mobility-related signaling for each attached MN, and performs mobility management on behalf of MNs by recording their movements. In a PMIPv6 domain, there are multiple MAGs, and AN is a network composed of link-layer access devices such as AP or BS which connects to MAG.

FPMIPv6 adopts the similar idea of FMIPv6 which sets up a bi-directional tunnel between PMAG and NMAG to tunnel the packets during the handover procedure.

On basis of FMIPv6 [7], FPMIPv6 extends Handover Initiate (HI) and Handover Acknowledge (HAck) messages to carry MN's Network Access Identifier (NAI), Home Network Prefix (HNP) and IPv4 Home Address from PMAG. However, for Router Solicitation, PIPv6 does not apply "Proxy Advertisement" (RtSolPr), the "Proxy Router Advertisement" (PrRtAdv), "Fast Binding Update" (FBU), "Fast Binding Acknowledgment" (FBack), and the "Unsolicited Neighbor Advertisement" (UNA)and for that the MN is not directly involved in FPMIPv6, and these messages, therefore, will be ignored by MAGs.

### 2.1.1FPMIPv6Operation Modes

FPMIPv6 has two operation modes based on the setup timing of bi-directional tunnel between PMAG and NMAG: the predictive mode in which the tunnel is established prior to attachment of the MN to the NMAG, and the reactive mode in which tunnel is established after the MN attaches to the NMAG. In both models, all the MAGs have to buffer the packets during the detachment procedure.

Figure 2 shows the operation flow of P-FMIPv6 in a predictive mode which is also called the PMAG-initiated handover.

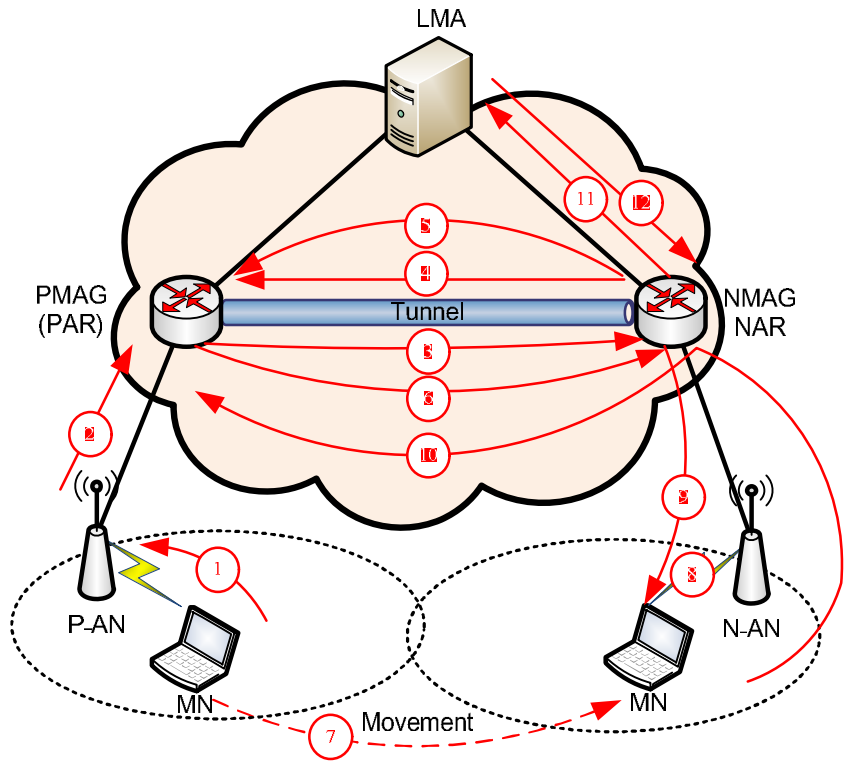

Figure 2 The predictive fast handover flow

The detailed operation flow is explained below:

(1) MN detects an imminent handover and reports its ID and New AP ID (Target AP) by access technology specific methods.

(2) P-AN sends a handover indication including MN ID and New AP ID to PMAG based on access technology specific methods.

(3) PMAG sends an HI message to NMAG with 'P' flag including MN ID, HNP(s) and LMA address, and MN link-layer ID such as MN LL-ID and MN LLA-IID to initiate the pre-registration procedure.

(4) NMAG replies with a HAck message to PMAG with 'P' flag and code value that indicates the pre-handover result.

(5) NMAG sends the HI message with 'U' or 'F' flag to optionally request the PMAG to buffer or forward packet at a later and appropriate time, respectively.

(6) PMAG will set up a bi-directional tunnel between PMAG and NMAG with Previous CoA and New CoA, and forwards the packets destined for MN if it receives an HI message with ‘F’. These packets will be buffered at NMAG before MN is attached.

(7) MN performs the handover from P-AN to N-AN based on the specific access technology operations.

(8) MN sets up a physical-layer connect with the N-AN and NMAG, and configures IPv6 address.

(9) NMAG forwards the packets to MN via N-AN.

(10) MN sends uplink packets to NMAG and NMAG forwards them to PMAG, while NMAG sends downlink packets to MN directly.

(11) NMAG updates the binding cache by sending PBU to LMA.

(12) If the binding is a success, LMA updates the tunnel between LMA and serving MAG via PBA, and all the packets will be forwarded via NMAG.

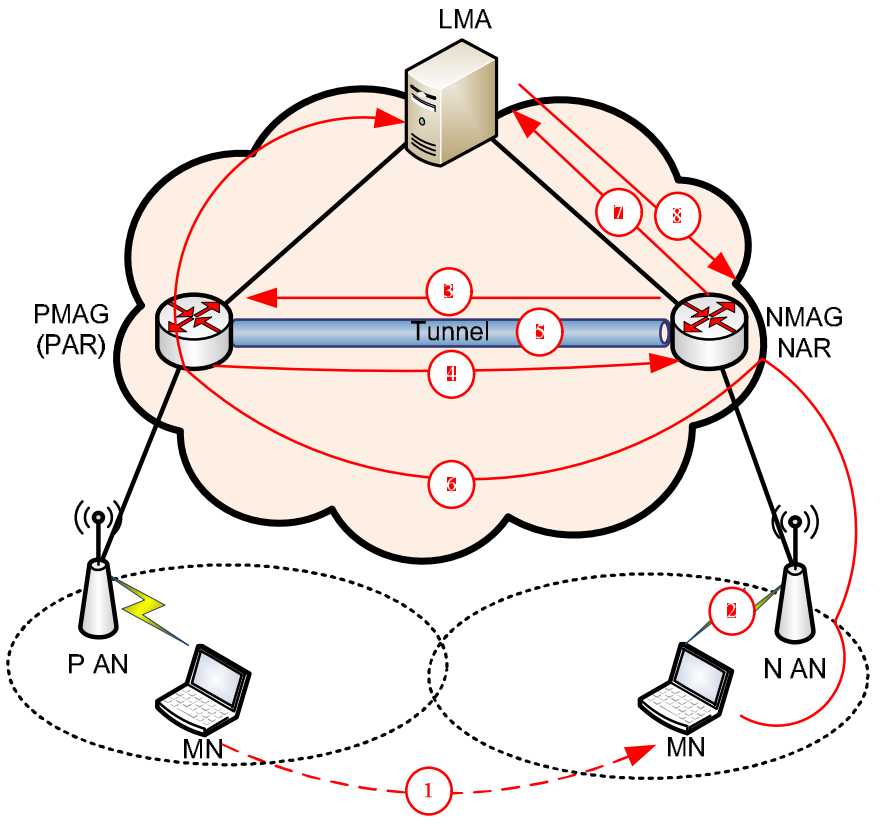

Figure 3 The reactive fast handover flow

Figure 3 shows the reactive mode of P-FMIPv6 which is also called the NMAG-initiated handover.

(1) At first, the MN handovers from P-AN to N-AN.

(2) MN establishes a new connection with N-AN, and transfers the MN ID and old AP ID to NMAG to identify the PMAG (substituted for UNA and FBU).

(3) NMAG sends an HI message with ‘P’ flag including MN ID to PMAG. If NMAG wants to set up a tunnel between PMAG and NMAG, it should also set ‘F’ flag in HI.

(4) PMAG replies with a HAck message to NMAG with ‘P’ flag including HNP(s), IPv4-MN-HoA, MN LLID (optional), the address of LMA and the other information requested by NMAG.

(5) If the HI message sent by NMAG is with ‘F’ flag, a bi-directional tunnel between PMAG and NMAG will be set up, and used to forward the packets destined for MN.

(6) MN sends uplink packets to NAMG via N-AN, and forwards to PMAG, and finally deliveries to LMA.

(7) NMAG updates the related binding cache by sending PBU to LMA.

(8) LMA updates the tunnel between LMA and serving MAG by reply a PBA, and all the packets will be forwarded via NMAG.

### 2.1.2 FPMIPv6Problems

FPMIPv6 is a network layer mobility support solution which provides a fast handover interaction framework and defines the related signaling messages format to reduce the handover delay and packets loss. However, to implement and deploy FPMIPv6 in large scale, it should consider the link layer operations and specific access technology. Therefore, it still has the following problems.

(1) Lack of definition of handover triggers events. For example, FPMIPv6 just gives a report message to notify the imminent handover in PMAG-initiated mode, which does not provide the operation in detail.

(2) Lack of candidate network discovery and selection mechanism which may result in the handover failure.

(3) Lack of handover execution procedure and link-layer specific operations in detail.

(4) Lack of detailed explicit heterogeneous handover mechanism.

Due to these problems, FPMIPv6 should incorporate other mechanisms such as MIH to support heterogeneous handover.

## 2.2 Overview of MIH

IEEE 802.21 (MIH) [11] aims to realize heterogeneous handover without service disruption, which can be a complementation of FPMIPv6. MIH contains handover initiation and preparation to search new link and set up a new link. The adaptation of MIH can facilitate service and maximize handover efficiency by combining with upper layer mobility management solutions.

MIH maintains a global network map which records available networks such as 802.11/16/22, 3G/4G, and their link layer information including neighbor maps, and higher layer services. More specifically, MIH locates between the layer 2 and the layer 3, and it abstracts the information exchange procedure of different link layer protocols. It also defines a unified interface Service Access Points (SAP) to provide services for upper layer by hiding the heterogeneity of the different link layer protocols in terms of topological and location information.

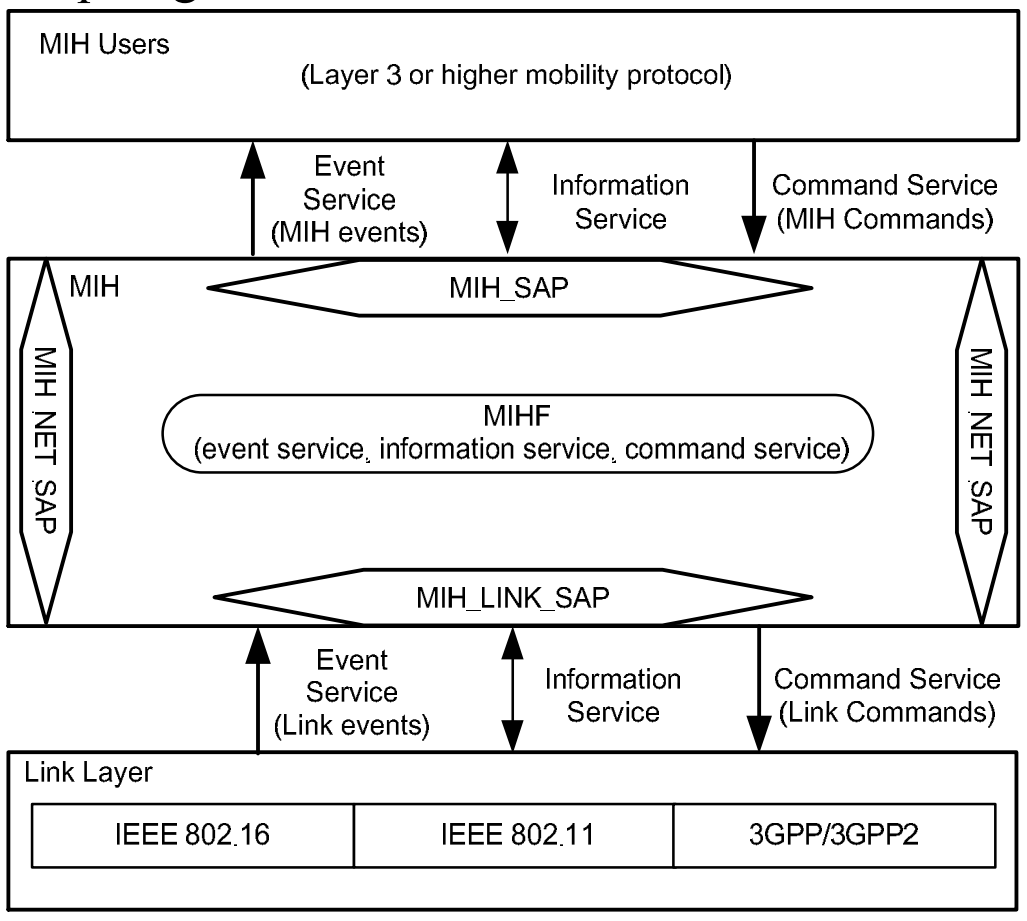

Figure 4 The MIH frameworkmodel [11]

Figure 4 shows the typical MIH framework which consists of MIH users such as mobility management protocols, a mobile node such as a smartphone, and networks such as 3GPP/3GPP2. All information exchanges are through SAPs, which have three types: a)

The MIH_SAP- which is the interface between MIHF and its users. It provides sharing of MIHF-generated events.It also communicates the link-layer events. b) The MIH_LINK_SAP- which is an abstract media dependent interface between the MIHF and lower layer protocols stacks and it provides the media-specific SAPs for different link-layer technologies. c) The MIH_NET_SAP- which is an abstract media dependent interface of the MIHF responsible for transport services over the data plane. It handles the information and messages exchanges with remote MIHFs.

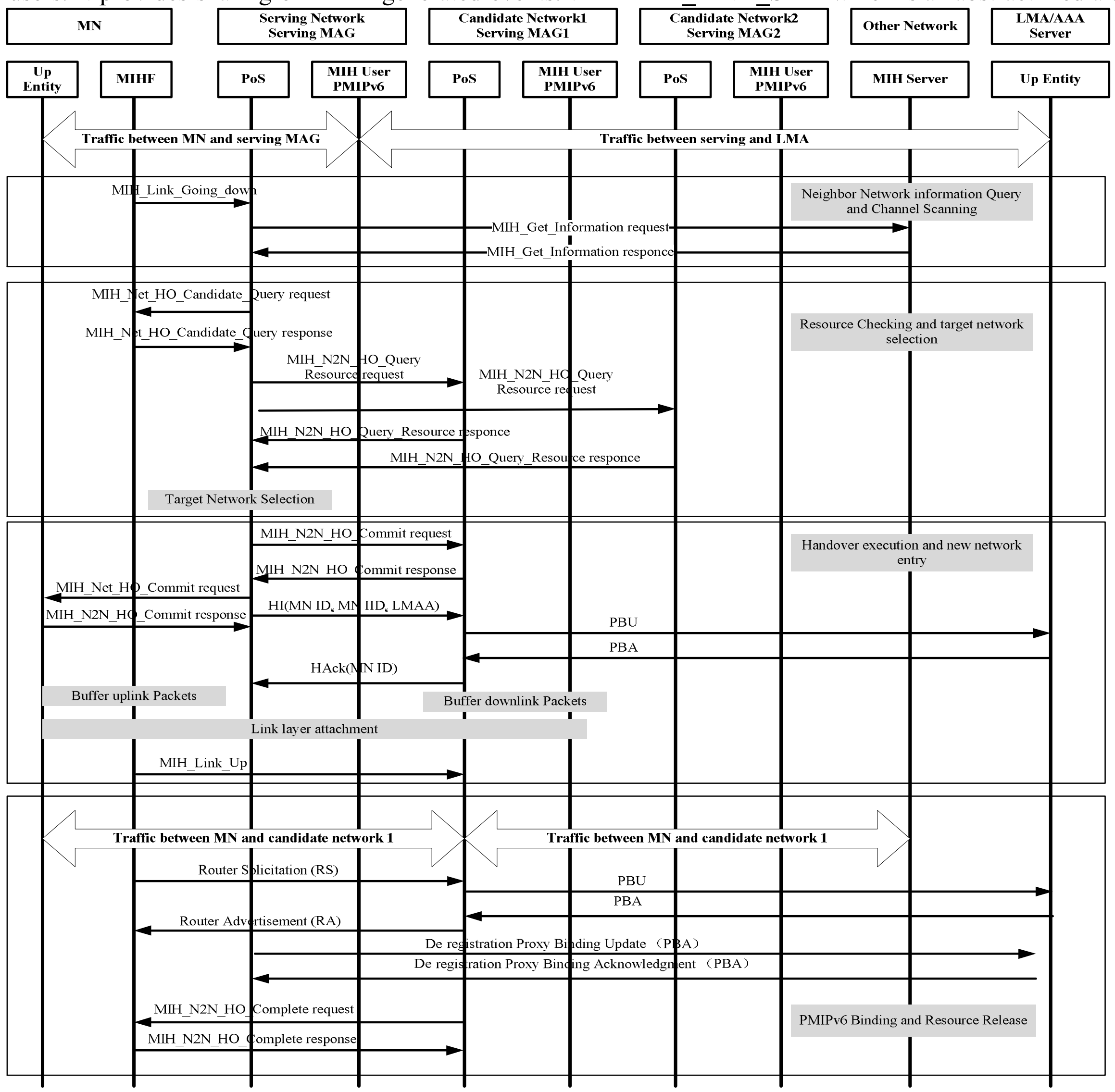

Figure 5 PMIPv6 assisted MIH using fast handover procedure [12] [19] [21]

### 2.3 Related Work of Integration Models

IEEE 802.21 provides two integration schemes of MIH and PMIPv6, namely network-initiated handover and mobile-initiated handover, whose handover procedures generally consist of information query, resource availability check, resource preparation, new L2 connection establishment, link up indication, IP connectivity restored, higher layer handover execution (PMIPv6 handover operations) and resource release. In the resource preparation phase, both of them introduce the pre-registration mechanism to reduce the handover delay and buffer the packets in target MAG. However, they have to acquire the MN's profile from AAA server or LMA, which may introduce unexpected delay. Besides, both of them are in faced with the packet disorder caused by buffer and larger handover delay problems due to lack of efficient L2 trigger mechanism. Therefore, several enhanced schemes were proposed, which are generally based on L2optimized mechanism, pre-authentication and pre-registration to realize fast handover.

Considering L2 scanning delay as one main component of handover delay, Kim *et al.* [19] proposed a low latency proactive handover scheme for PMIPv6 with MIH, which can reduce the overall L2 scanning time. In fact, this reduction of L2 scanning time is mainly specific to Wi-Fi based on MIHF which provides the channel configuration information of each AP, so that MN can only scan the configured channels in each AP, not for all channels [20]. However, this scheme requires the buffering packets from both the LMA and the nMAG, which may result in the out-of-order problem. Therefore, they improved it by transient binding [21] and modified pre-registration procedure. After that, to provide the QoS guaranteed real-time services, Kim et al [9], optimized predictive PFMIPv6 handover scheme which supports the pre-registration via MIH_Net_HO_Candidate_Query request and response messages, and supports the pre-authentication with EAP. At the same time, to reduce the handover delay, it adopts the prediction-based smart channel scanning [22]. Figure 5 shows its handover procedure [12] [19] [21] (In the following part, we note this scheme as fast handover solution).

The detailed operation flow is presented below:

(1) Once an MN detects the link going down (it detects the signal strength becoming weak), it sends MIH_Link_Going_down message to its serving network to search the available neighbor networks. The serving network queries the related information from MIIS server through the MIH_Get_Information request message and MIH_Get_Information reply message.

(2) Resource checking and target network selection stage:

Serving MAG and MN negotiates the candidate MAGs through MIH_Net_HO_Candidate_Query request and response messages, and serving MAG checks the resources availability of all the candidate MAGs by MIN_N2N_HO_Candidate_Query request and response messages. By this way, the target MAG is selected.

(3) Handover execution and new network entry stage:

The serving MAG sends MIH_N2N_HO_Commit request message to request the target network to allocate resources, and then serving MAG and target MAG use the HI/HAck messages to transfer the context information including MN ID, MN IID, LMAA, and perform the pre-registration. Once the target network received the HI message, it will set up the transient BCE via PBU/PBA with a transient binding option to simultaneous reception from both networks during handover. After that, serving MAG informs MN about the target network information and indicates the completion of pre-authentication.

(4) PMIPv6 binding and resource release:

MN sends the MIH_Link_Up message to the target network to capture the buffered packets. The target MAG and LMA complete the binding update via PBU and PBA. After that, the serving network will tear down the bi-directional tunnel to LMA via De-registration PBU/PBA, and finally, the target MAG sends the MIH_N2N_HO_Complete to serving MAG to release the resource.

This fast handover scheme can improve the performance of heterogeneous handover, however, it still has the following problems: 1) It is a mobile-initiated method for that it introduces five signals between MN and serving MAG, which may increase the power consumption on MN and affects its execution effectiveness. Also, it does not comply with the design principle of PMIPv6 that reduces the involvements of MN in mobility management. 2) It does not distinguish MN's network interfaces which make it difficult to describe cooperation mechanism among different network interfaces, and unclear to support heterogeneous handover. 3) It is mainly used for 802.11, which can be further extended to a universal network by IE_POA_CHANNEL_RANGE in MIIS.

The other solution such as PMIPv6 assisted MIH using MIHF at network side only scheme [17] is dependent on the network to exchange the signaling messages. In such case, it is difficult to support heterogeneous handover as the S-PoS does not know the wireless interfaces information of mobile nodes and it, therefore, cannot select the heterogeneous candidate networks. To make it clear, we summarized related existing schemes as shown in Table 1.

Table 1 The summary of existing PMIPv6 and MIH integration model

| Schemes / Metric | The PMIPv6 assisted MIH using standard approach | PMIPv6 assisted MIH using wireless modified MIHF signals | PMIPv6 assisted MIH using MIHF at network side only | PMIPv6 assisted MIH with neighbor discovery | PMIPv6 assisted MIH with Handover Coordinator (HC) | PMIPv6 assisted MIH using fast handover scheme |
|---|---|---|---|---|---|---|
| **Triggermechanism** | MN's MIHF | No | L2 trigger | MN trigger | MN trigger | MN trigger |
| **Initiation method** | MN | Network | Network | undefined | HC | Network |
| **Candidate Network provider** | MN | S-PoS | S-PoS | undefined | S-PoS | MN |
| **Resource check point** | S-PoS | S-PoS | S-PoS | No | Undefined, maybe is HC | S-PoS |
| **target network selector** | MN | S-PoS | S-PoS | undefined | S-PoS | S-PoS |
| **Packet buffer point** | LMA | LMA | LMA | S-PoS | Undefined | C-PoS (DL)<br>MN (UL) |
| features | (1) large power comsumption for that MN is involved in too many signaling exchanges;<br>(2) large handover delay due to the lack of the fast handover support. | (1) Lack of the trigger mechanism;<br>(2) S-PoS selects the target network which is difficult to support trigger;<br>(3) lack of fast handover;<br>(4) larget packet during the handover | (1) Network side selects target network which is difficult to support heterogeneous handover;<br>(2) Lots of information is handled by S-PoS which can reduce the function requirement of MN | (1) The function of ND has been involved in MIH;<br>(2) unclear how to select the target network;<br>(3) lack of the resource check. | (1) Support thepre-resgristation and pre-authentication;<br>(2) Bi-casting can reduce packets loss;<br>(3) Lack of thedetailed HC selection mechanism, and the resource check point is unclear | (1) Adopt Transist Binding mechanism which is difficult to handle with the disorder;<br>(2) Only focus on single interface, lack of the analysis of multiple interfaces. |

Table 2 The parameters of extended MIH_N2N_HO_Commit request message

| Parameter | Type&Value | Data type | Description |
|---|---|---|---|
| MN LLA-IID | MN LLAID (Extended) Value=101 | LINK ADDR or Interface identifier | It is MN link-local address interface identifier which is used to identify the attached interface of a MN, and it can be MAC address, cell ID or other link address |
| LMAA | LMA Address (Extended) Value=102 | TRANSPORT_ADDR | Identify the address of LMA which can be IPv4 or IPv6. |
| HNP | Home Network Prefix (Extended) Value=103 | LIST(HNP)where HNP=SEQUENCE( UNSIGNED_INT(1) OCTET_STRING(16)) | Identify the list of MN's IPv6 home network prefix(es) assigned to the Target MN link. UNSIGENED_INT: IP_PREFIX_LEN |

In this paper, we extend these operations and integrate FPMIPv6 with MIH. Considering that the link layer support is beneficial to the predictive fast handover mode, we mainly focus on the integration scheme of predictive mode and MIH.

# 3 The Proposed Solution

Compared with PMIPv6, FPMIPv6 optimizes the handover procedure by L2 trigger and pre-registration. In the proposed integration model, it performs the MIH discovery procedure to find an MIHF's capabilities of MIH services through the standard protocol or media-specific mechanisms (i.e., IEEE 802.11 Beacon frames, IEEE 802.16 downlink channel descriptor (DCD), IEEE 802.11management frames, or IEEE 802.16 management messages).

The main idea is to extend the MIH messages to carry the required information for triggering the pre-registration procedure without the additional MIH user signaling messages, such as HI/HACK. More specifically, we extend MIH_N2N_HO_Commit request message to carry MN ID, MN LLA-IID, LMAA and HNP to trigger the binding between nMAG, and extend MIH_N2N_HO_Commit response message to indicate the handover status. Besides, the transient binding is introduced which will be initiated by LMA to support transient BCE for both pMAG and nMAG to reduce the packet loss.

## 3.1 MIH message extensions

### 3.1.1 Extension of MIH_N2N_HO_Commit request

According to [18] Annex L, the TLV code has been assigned to 100, so the parameters of extended MIH_N2N_HO_Commit request message are shown in Table 2. In this message, we extended three parameters as shown in following.

(1) MN LLA-IID

We define a new TLV value as 101 to represent MN's link-local address interface identifier, which is used by MAG to associate PMIPv6 tunnel with the access link that MN attached in case of the point-to-point link.

(2) LMA Address

This extended TLV with value 102 carries the address of LMA that can be IPv6 or IPv4.

(3) Home Network Prefix

It is used to identify the list of MN's IPv6 home network prefix (es) assigned by LMA to the MN's target link. This is a list of HNPs which consists of 1 octet to identify the length of HNPs and HNPs.

The extended message format of MIH_N2N_HO_Commit request is shown in Figure 6. By introducing these fields, the MN's profile information can be delivered to the potential target networks to perform the pre-register and set up a tunnel.

### 3.1.2Extension of MIH_N2N_HO_Commit response

This message is extended to include the handover statuses that are defined by IANA. The current statuses defined by MIH only consist of success (value=0), unspecified failure (value=1), rejected (value=2), authorization failure (value=3) and network error (value=4), while others are undefined (value = 5~255). These statuses is simpler compared with the IANA, therefore in the proposed scheme, we extended the existing status code by adopting the status codes defined by IANA which are shown in Table 3.

Figure 7 shows the format of an extended MIH_N2N_HO_Commit response message. By using these extensions, the HI/Hack can be replaced by MIH_N2N_HO_Commit messages. Besides, by introducing the transient binding initiated by LMA, the packet of downlink packets can be transmitted to pMAG and nMAG to reduce the packet loss. Besides, there is no tunnel, so the traffic overhead will be reduced. Furthermore, our solutions can also be improved by using the optimized L2 mechanism when applied in the specific wireless network types.

## 3.2The proposed scheme operation flow

The proposed integration model is based on the following assumptions:

(1) All the network entities support MIH function;

(2) MIIS server is already set up and records the neighboring network information in advance;

(3) All the MIH entities have been configured and registered;

(4) MN has multiple network interfaces, and each network interface can work at the same time, but only one interface is used to transmit the packets. This assumption considers that an MN is energy-limited and the use of multiple network interfaces may reduce its available time although it can increase the throughput.

| 4 | 1 | 1 | 1 | 1 | 7 | 1 | 4 | 2 | 12 |
|---|---|---|---|---|---|---|---|---|---|
| VER | Ack Req (1) | Ack Rsq (0) | UIR (1) | M | FN | Rsvd1 (0) | Service identifier SID (3) | Opcode (1) | Action identifier (AID) (9) |
| Rsvd2 (0) | Transaction ID | | | | | | Variable Payload Length | | |

| | | |
|---|---|---|
| Type=1 | Length | Data type=MIHF ID (variable payload length, maximum length is 253octets ) |
| Type=2 | Length | Data type=MIHF ID (variable payload length, maximum length is 253octets ) |
| Type=52 | Length | Data type=MIHF ID (variable payload length, maximum length is 253octets ) |
| Type=59 | Length | Data type=LINK ID (1 octet link type, and variable payload length for various link addresses) |
| Type=60 | Length | Data type=LINK ADDR (variable payload length for various links ) |
| Type=Interface Identifier | Length | Data type=Interface Identifier (IEEE EU-64, RFC 4291) |
| Type=LMA Address | Length | Data type=IP_ADDR (2 octets for address type, variable length address) |
| Type=HNP | Length | Data type=(prefix length, IP6) (1 octet for prefix length, 16 octets for IPv6 address) |

Figure 6 The message format of extended MIH_N2N_HO_Commit request

| 4 | 1 | 1 | 1 | 1 | 7 | 1 | 4 | 2 | 12 |
|---|---|---|---|---|---|---|---|---|---|
| VER | Ack Req (1) | Ack Rsq (0) | UIR (1) | M | FN | Rsvd1 (0) | Service identifier SID (3) | Operation code Opcode (1) | Action identifier (AID) (9) |
| Rsvd2 (0) | Transaction ID | | | | | | Variable Payload Length | | |

| | | |
|---|---|---|
| Type=1 | Length | Data type=MIHF ID (variable payload length, maximum length is 253octets ) |
| Type=2 | Length | Data type=MIHF ID (variable payload length, maximum length is 253octets ) |
| Type=3 | Length | Data type=Status (1 octet, 0=255 ) |
| Type=52 | Length | Data type=MIHF ID (variable payload length, maximum length is 253octets ) |
| Type=59 | Length | Data type=LINK ID (1 octet link type, and variable payload length for various link addresses) |
| Type=57 | Length | Data type=ASGN_RES_SET (Assigned resource information) |

Figure 7 The format of extended MIH_N2N_HO_Commit response message

Table 3 Comparison of status code defined in MIH and IANA [7][17]

| Status code | MIH definition | IANA definition |
|---|---|---|
| 0 | success | Handover accept or success |
| 1 | Unspecified Failure | Handover Accepted, NCoA not valid [RFC5568] |
| 2 | Rejected | Handover Accepted, NCoA assigned [RFC5568] |
| 3 | Authorization failure | Handover Accepted, use PCoA [RFC5568] |
| 4 | Network error | Message sent unsolicited [RFC5568] |
| 5 | Unassigned | Context Transfer Accepted or Successful [RFC5949] |
| 6 | Unassigned | All available Context Transferred [RFC5949] |
| 7-127 | Unassigned | Unassigned |
| 128 | Unassigned | Handover Not Accepted, reason unspecified [RFC5568] |
| 129 | Unassigned | Administratively prohibited [RFC5568] |
| 130 | Unassigned | Insufficient resources [RFC5568] |
| 131 | Unassigned | Requested Context Not Available [RFC5949] |
| 132 | Unassigned | Forwarding Not Available [RFC5949] |
| 133-255 | Unassigned | Unassigned |

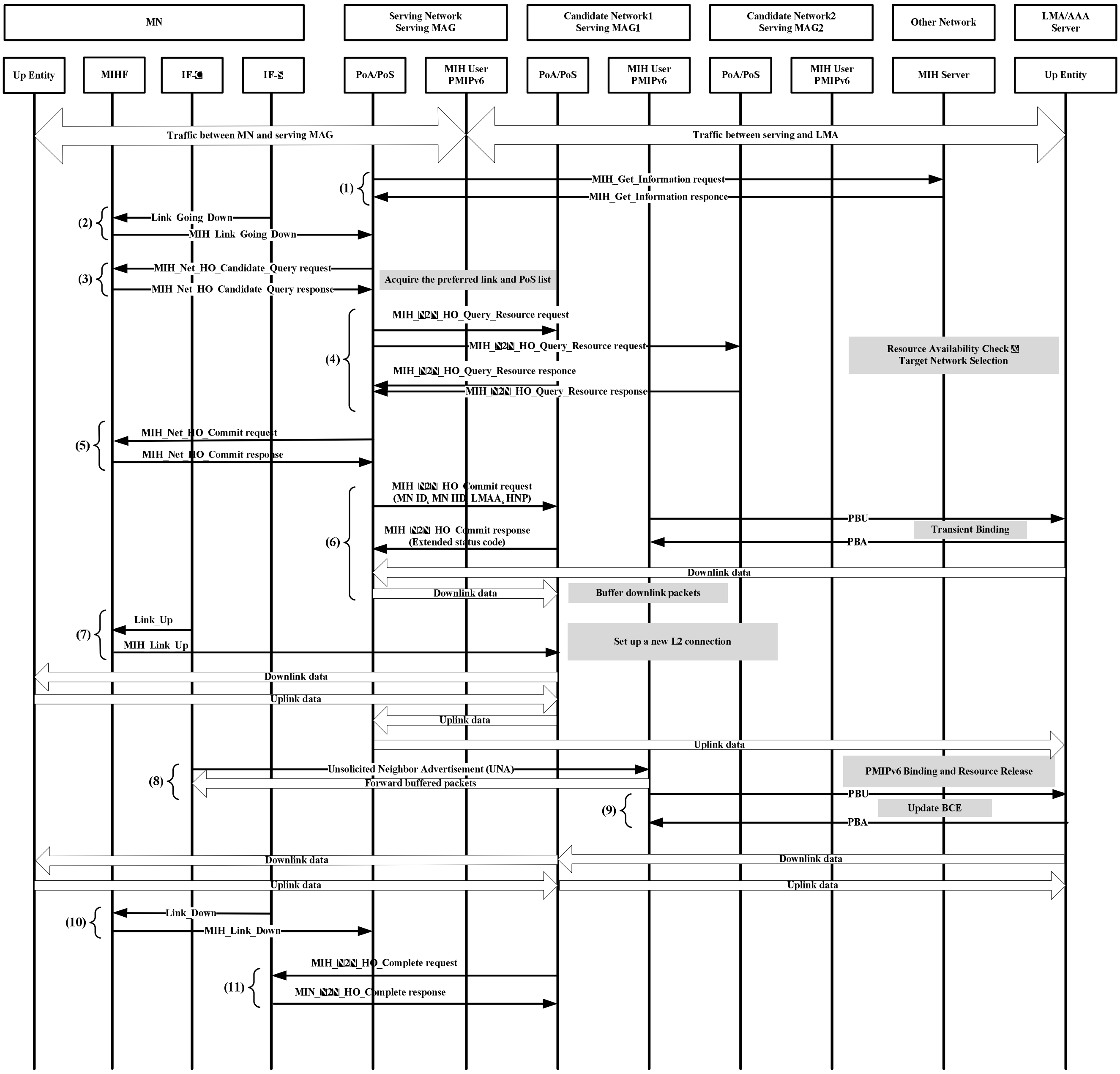

Figure 8 The proposed FPMIPv6 and MIH integration model operation flow

Figure 8 shows the proposed integration model, in which IF-S notifies the current serving interface, and the IF-C notifies the candidate network interface.

Assuming that each interface can detect their link status independently, and MN is equipped with two kinds of interfaces called serving interface (IF-S) and candidate interface (IF-C). These network interfaces can be homogeneous or heterogeneous. In the following section, the serving network is, in fact, the pMAG of PMIPv6, while the target network is the nMAG.

(1) In bootstrap stage, each PoA/PoS acquires its neighboring network information through MIH Information Server via MIH_Get_Information request and response messages to facilitate seamless handover. These neighboring networks contain not only the homogeneous access networks but also the heterogeneous access networks. The serving network will update its neighboring network information once an MN is attached or periodically refreshed.

(2) The IF-S of MN detects the link status and sends an MIH_Link_Going_down indication message to its serving network once it finds the current link quality is going down.

(3) The serving network sends the MIH_Net_HO_Candidate_Query request message to the MN for acquiring the candidate networks, and MN replies MIH_Net_HO_Candidate_Query response message that carries the MN's preferred link and PoS list in sequence.

(4) After receiving the response message, the serving network requires the resource preparation to the networks in the list (candidate networks) by MIH_N2N_HO_Query_Resource request and response messages.

(5) After that, the serving network will select the target network based on the available resource of candidate networks and sends extended MIH_N2N_HO_Commit request/response messages to the candidate network to prepare the resource. By this operation, the MN's related profile information will be exchanged between serving network and the target network. At the same time, the target network will send PBU message with a transient flag to LMA to set up the transient BCE. During the handover, the downlink packets will be buffered in the target network via LMA and serving network to reduce the packet loss.

(6) Once the resource is successfully prepared, the serving network sends the MIH_Net_HO_Commit request message to command MN for handover to the target network.

(7) The MIHF will send an MIH_Link_Up indication message to IF-C to set up the L2 connection and replies an MIH_Net_HO_Commit response message to the serving network.

(8) After that, MN sends an Unsolicited Neighbor Advertisement (UNA) message to the target network. Once target MAG receives this message, the link layer and IP layer will be established. At the same time, this message will trigger target MAG to forward buffered packets to MN.

(9) Once the IP layer is established, target MAG and LMA will perform the PBU/PBA procedure to update the BCE and bi-directional tunnel between target network and LMA. After updating the BCE, the target network will become the serving network and forwards the native packets to MN.

(10) Once MN gets the native packets from the new network, it will send the MIH_Link_Down to the old serving network to tear up the original binding and release it.

(11) After that, the serving network sends the MIH_N2N_HO_Complete request/response to release the resource.

# 4 Performance Evaluations

In our performance analysis, we compare our solution with the standard handover scheme and the fast handover solution in terms of handover delay and signaling cost.

To simplify the analysis, we note the distance between X and Y as $H_{x\text{-}y}$ which equals to $H_{y\text{-}x}$. The delay between X and Y is denoted as $D_{x\text{-}y}$. Besides, the signaling and delay between PoA and PoS are neglected. The related parameter setting is shown in

Table 4, and some values of them are based on [23][24].

The signaling cost is computed by *hop*message_size.* The typical MIH message size for the Events and Commands Services ranges between 50 to 100 bytes, and therefore, the transport of MIH message should solve the message fragmentation of UDP and the message concatenation problems. In the analysis, the message size is subjected to the numbers of neighboring networks, candidate networks, HNPs and so on. To simplify the notations, we adopt the abbreviations specified in Table 5. Besides, to make the analytical results more realistic, the wireless link failure probability ($P_f$)is considered.

Table 4 The parameters list

| Symbol | Value(s) | Description |
|---|---|---|
| a | 36000m | City section length |
| b | 24000m | City section width |
| $R$ | 100m | Cell radius |
| $V$ | 1~50 m/s | Average speed of MN |
| $A$ | 1 | Unit transmission cost over wired link |
| $B$ | 1.5 | Unit transmission cost over wireless link |
| $N$ | 6 | Average number of neighboring networks |
| $m$ | 6 | Average number of preferred PoAs |
| $P_f$ | 0.5 | Wireless link failure probability |
| $\tau$ | 20ms | Interframe time |
| $\rho_f$ | 0.1 | Frame error rate (FER) over wireless link |
| $d$ | 10m | Distance between adjacent roads |
| $T_{max}$ | 70s | Maximum pause time in a location |
| $T_{L2}$ | 45.35ms | The link layer handover delay |
| $H_{MN\text{-}MAG}$ | 1 hop | Average distance between MN and MAG |
| $H_{MAG\text{-}LMA}$ | 10 hops | Average distance between MAG and LMA |
| $H_{MAG\text{-}MIIS}$ | 10 hops | Average distance between MAG and MIIS |
| $H_{MAG\text{-}MAG}$ | 10 hops | Average distance between MAG and MAG |

**4.1 Handover delay analysis**

The handover delay is the time interval between the moments when an MN loses connectivity with its serving MAG until the moment it receives the first packets from the target MAG. For the pre-registration mechanisms, the first packet can be the buffered packet.

To evaluate handover delay, we adopt the method in [24]. Assuming that $\tau$is interframe time, $\rho_f$ is the frame error rate over the wireless link, $L_p$ and $L_f$ are the packet size and frame size, respectively, and $D_{wl}$ is the wireless link delay mainly depending on the L2 technology being used.

The one-way packet transport delay over the wireless link $d_{wl}(L_p)$ can be expressed as

$$d_{wl}(L_p) = d_{frame} + (k-1)\tau \qquad (1)$$

Where $k = \lceil L_p / L_f \rceil$ , and $d_{frame}$ is the one-way frame transport delay of wireless link, which can be expressed as:

$$d_{frame} = D_{wl}(1-\rho_f) + \sum_{i=1}^{n}\sum_{j=1}^{i} p_{i,j}(2i \times D_{wl} + 2(j-1)\tau) \qquad (2)$$

Where $p_{i,j}$ is given as:

$$p_{i,j} = \rho_f (1-\rho_f)^2 ((2-\rho_f)\rho_f)^{((i^2-i)/2+j-1)} \qquad (3)$$

Assuming that $BW_{wired}$and $D_{wired}$ are the bandwidth and the latency of wired links, respectively, the one-way packet transportation delay over a wired link through $h$ hops can be expressed as

$$d_{wd}(L_p, \mathrm{h}) = \frac{L_p \times h}{BW_{wired}} + D_{wired} \qquad (4)$$

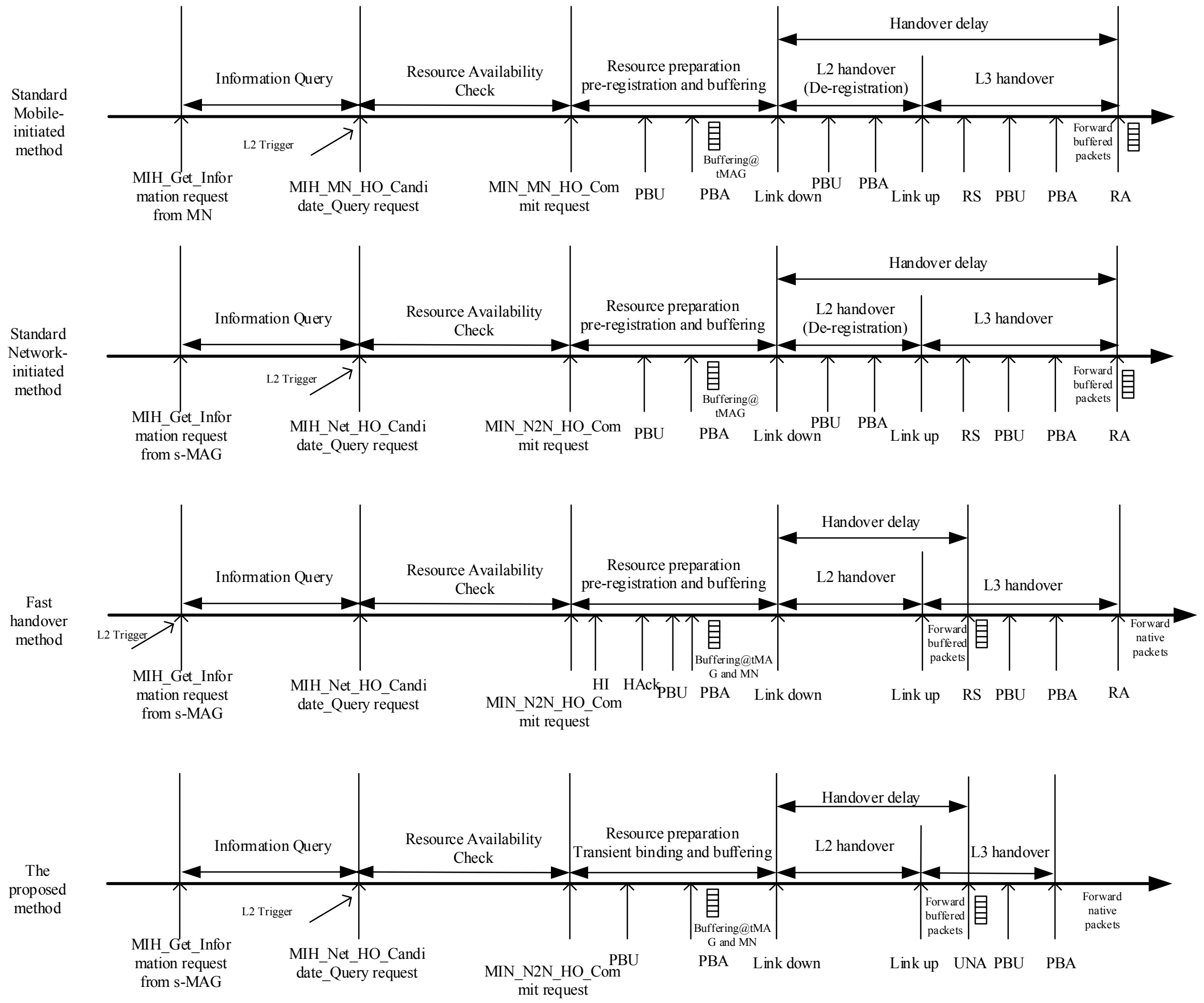

Figure 9 The timing diagram of handover procedures

Figure 9 shows the handover delay of standard handover procedure (mobile-initiated and network-initiated method), fast handover and the proposed solution. From Figure 9, it can be noticed that the fast handover and the proposed method have the similar handover delay.

Based on Figure 9, the handover delay of mobile-initiated method and network-initiated method of standard handover scheme can be expressed as

$$t_{SH} = d_{wl}(M_{RS}) + d_{wl}(M_{RA}) + d_{wd}(M_{PBU}, H_{MAG_LMA}) + d_{wd}(M_{PBA}, H_{MAG_LMA}) + d_{wl}(L_D) + T_{L_2} \qquad (5)$$

Similarly, the handover delay of fast handover solution is shown in (6). Since the packets are buffered in target MAG, it should include the additional tunneling header which is 40 bytes.

$$t_{FH} = T_{L_2} + d_{wl}(M_{RS}) + d_{wl}(L_D + 40) \qquad (6)$$

The proposed solution can be expressed as

$$t_{OH} = T_{L_2} + d_{wl}(M_{UNA}) + d_{wl}(L_D + 40) \qquad (7)$$

Figure 10shows the handover delay versus different frame error rates. It is obviously that the increasing of frame error rate increases the handover due to retransmission. However, the handover delay of the proposed solution is less than others.

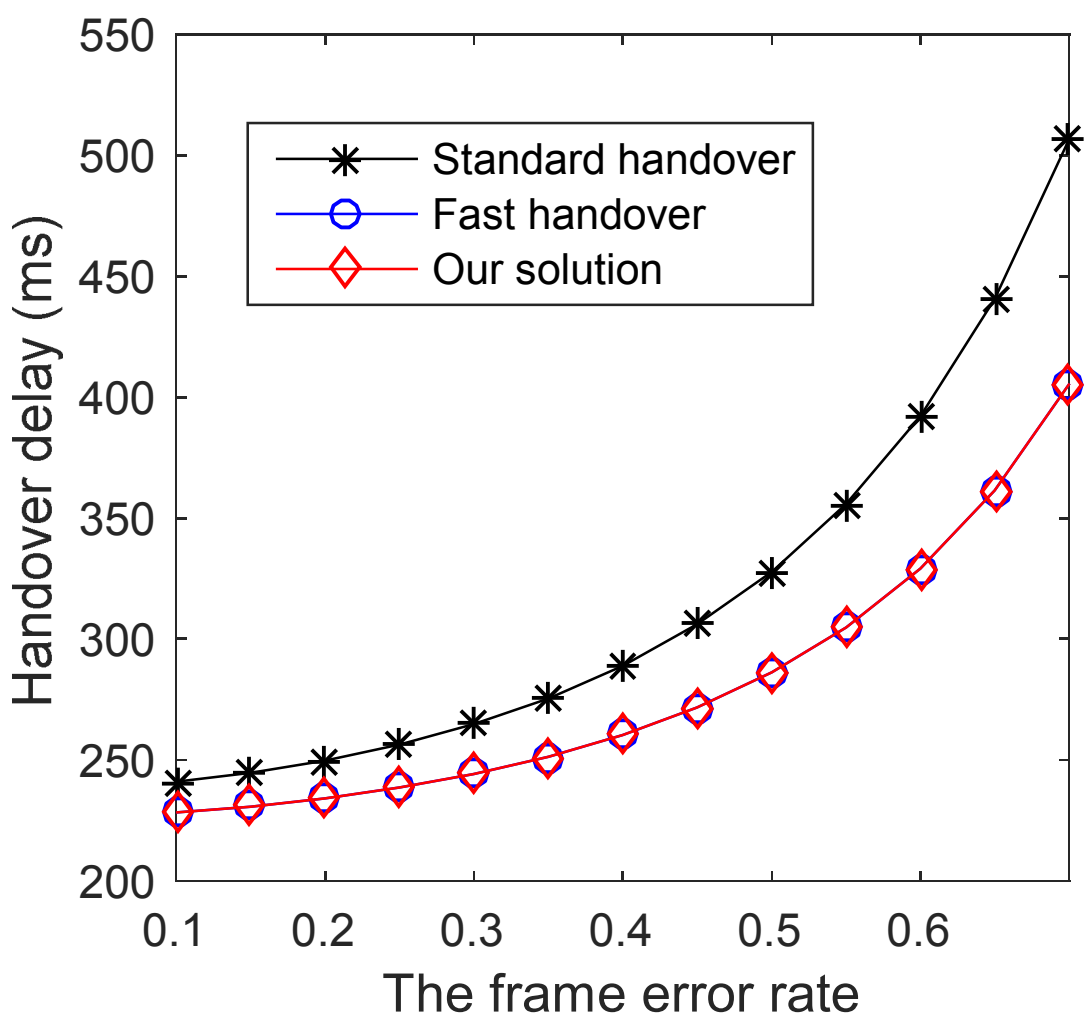

Figure 10 Handover delay versus the frame error rate

Figure 11 shows the handover delay versus the wireless layer delay. The increasing wireless layer delay results in the larger handover delay. Similar to Figure 10, the handover delay of the proposed method is less than the existing solutions.

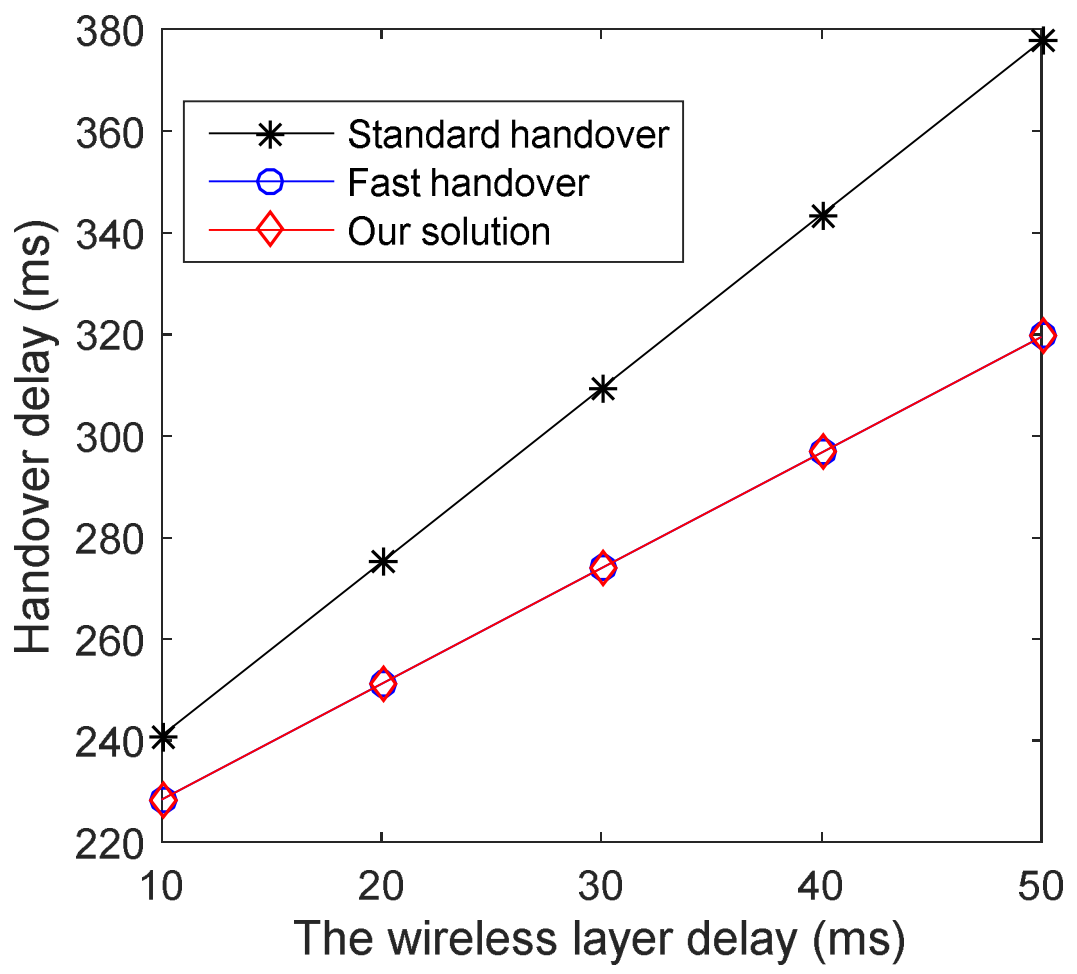

Figure 11 Handover delay versus wireless layer delay

## 4.2 Signaling cost analysis

In this paper, we adopt the city mobility model [25][27] to analyze the performance of the proposed scheme. In this model, mobile node moves in an epoch-based pattern, in which it starts at a defined point on the street, and then it randomly selects a destination. Once it reaches the destination, it will pause for a random time.

The city is supposed to be a rectangle of a*b. Let $d_x$ and $d_y$ represent the distance between adjacent horizontal roads and vertical roads, respectively. Both of them can be used to reflect the road density.

According to [25][27], the expected epoch length can be expressed as

$$E(L) = \frac{d_x(N_h+1)(N_h-1)}{3N_h} + \frac{d_y(N_v+1)(N_v-1)}{3N_v} \qquad (8)$$

Where $N_h = \lceil a/d_x \rceil$ and $N_v = \lceil b/d_y \rceil$

The expected number of subnet crossing can be expressed as

$$E(N_t) = E(N_x) + E(N_y) \qquad (9)$$

Where

$$E(N_x) = \frac{m(m+1)K_1}{6N_h^2}(6N_h - 4mK_1 + K_1 + 3) \qquad (10)$$

Where $K_1 = 2r/d_x$

$$E(N_y) = \frac{m(m+1)K_2}{6N_v^2}(6N_v - 4mK_2 + K_2 + 3) \qquad (11)$$

Where $K_2 = 2r/d_y$

Let $V_{min}$ and $V_{max}$ represent the minimum speed and maximum speed of mobile node, respectively. The expected time for an epoch is

$$E(T_t) = \frac{E(L)\times \ln|V_{\max}/V_{\min}|}{V_{\max} - V_{\min}} \qquad (12)$$

Assuming that mobile node in the destination follows the uniform distribution between [0, $T_{max}$], then the expected pause time $E(T_p)$ is calculated by 0.5*$T_{max}$.

Therefore, the number of handover per unit time can be calculated as follows.

$$E(N_c) = \frac{E(N_t)}{E(T_t) + 2E(T_p)} \qquad (13)$$

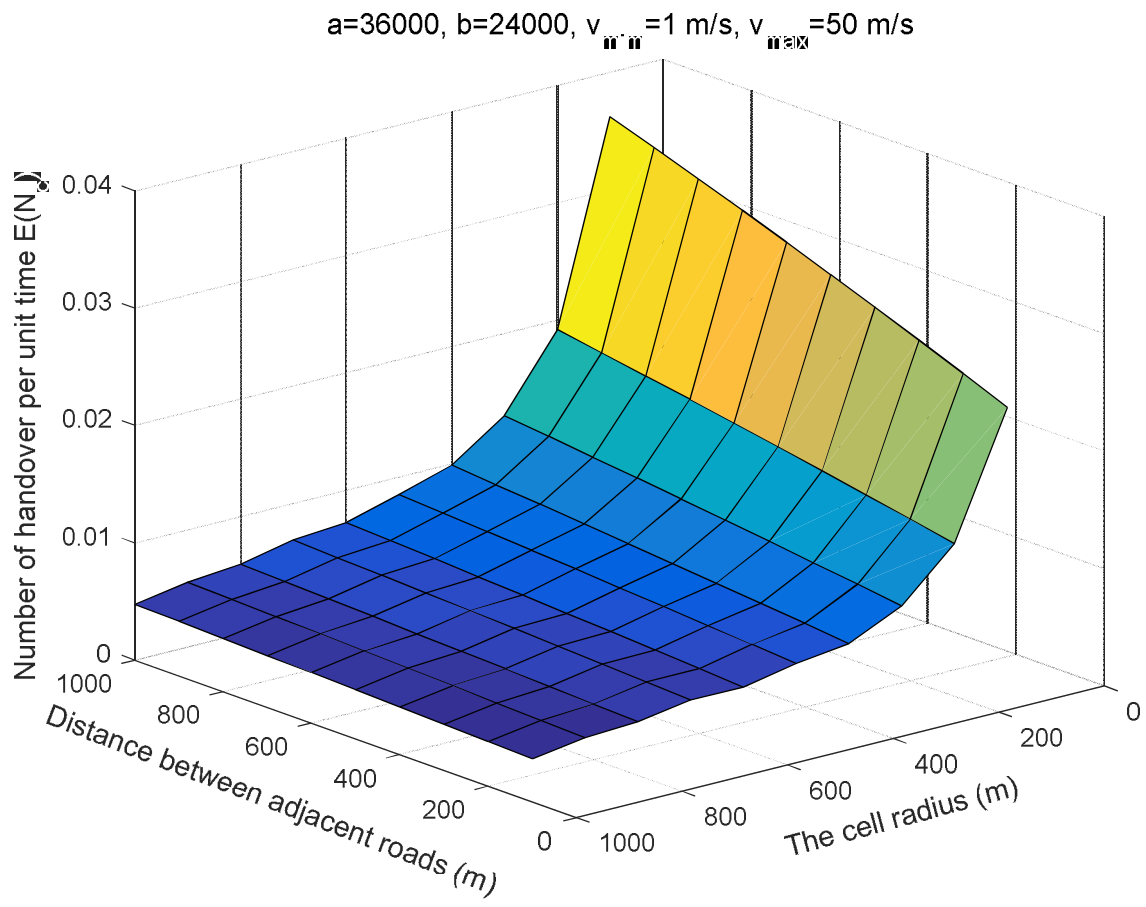

Figure 12 The number of handover per unit time versus cell radius and distance between adjacent road

Figure 12 depicts the number of handovers per unit time versus cell radius and distance between the adjacent roads. The city section is a rectangular area of 36000m*24000m, and the speed is set to [1, 50] m/s. The number of handovers increases as the distance between adjacent roads as well as the cell radius reduces.

Table 5 The sizes of messages used in the analysis

| Message name | Type | Size | Abbreviation |
|---|---|---|---|
| MIH_Link_Going_down | Event service | 78 | $M_1$ |
| MIH_Link_Up | Event service | 95 | $M_2$ |
| MIH_Get_Information request | Information service | 1500 | $M_3$ |
| MIH_Get_Information response | Information service | 1500 | $M_4$ |
| MIH_Net_HO_Candidate_Query resquest | Command service | 63+11*n+8*m*n | $M_5$ |
| MIH_Net_HO_Candidate_Query response | Command service | 77+101*m | $M_6$ |
| MIH_N2N_HO_Query Resource request | Command service | 150+11*m | $M_7$ |
| MIH_N2N_HO_Query Resource response | Command service | 165 | $M_8$ |
| MIH_N2N_HO_Commit request | Command service | 213 | $M_9$ |
| MIH_N2N_HO_Commit request (Extended) | Command service | 264 | $M_{9e}$ |
| MIH_N2N_HO_Commit response | Command service | 92 | $M_{10}$ |
| MIH_N2N_HO_Commit response (Extended) | Command service | 92 | $M_{10e}$ |
| MIH_Net_HO_Commit request | Command service | 122 | $M_{11}$ |
| MIH_Net_HO_Commit response | Command service | 103 | $M_{12}$ |
| MIH_N2N_HO_Complete request | Command service | 109 | $M_{13}$ |
| MIH_N2N_HO_Complete response | Command service | 112 | $M_{14}$ |
| MIH_MN_HO_Commit request | Command service | 75 | $M_{15}$ |
| MIH_MN_HO_Commit response | Command service | 78 | $M_{16}$ |
| AAA Query | Mobility management | 32 | $M_{17}$ |
| AAA Reply | Mobility management | 60 | $M_{18}$ |
| HI | Mobility management | 72 | $M_{HI}$ |
| HAck | Mobility management | 32 | $M_{HAck}$ |
| PBU | Mobility management | 76 | $M_{PBU}$ |
| PBA | Mobility management | 52 | $M_{PBA}$ |
| RS | Mobility management | 16 | $M_{RS}$ |
| RA | Mobility management | 64 | $M_{RA}$ |
| UNA | Mobility management | 52 | $M_{UNA}$ |

The messages sizes only consider the protocol header, while ignore the outer header such as IPv6 header.

Differenttothepreviousanalysiswhichassumesthatthesizeofallsignalingmessagesissame, we distinguish the sizes of different messages as shown in Table 5.

According to [26], the wireless link transmission fail probability is $P_f$, therefore the signaling cost of the given solution can be expressed as

$$S = E(N_c)*(P_f/(1-P_f)*S_{FH_WL}+S_{FH_W}) \quad (14)$$

Where $S_{FH_WL}$ means the signaling cost in the wireless link and $S_{FH_W}$ is the signaling cost in the wired link.

**(1) Standard handover solution**

For standard handover solution, the handover delay can be expressed as follows.

The messages between MN and serving MAG/target MAG ($H_{MN_MAG}$) includes: $M_3$, $M_4$, $M_5$, $M_6$, $M_{15}$, $M_{16}$, $M_{RS}$ and $M_{RA}$.

The messages between serving MAG and MIIS ($H_{MAG_MIIS}$) includes: $M_3$ and $M_4$.

The messages between serving MAG and candidate MAG ($H_{MAG_MAG}$) includes: $M_7$and $M_8$. In fact, there may be multiple candidate MAGs in the resource check phase. While in the handover delay analysis, we ignore the sending interval of $M_7$ and $M_8$.

The messages between serving MAG and target MAG ($H_{MAG_MAG}$) includes: $M_9$, $M_{10}$, $M_{13}$, $M_{14}$, $M_{PBU}$ and $M_{PBA}$.

The message between serving MAG and LMA ($H_{MAG_LMA}$) includes: AAA Query ($M_{17}$), AAA Reply ($M_{18}$), $M_{PBU}$ and $M_{PBA}$. AAA Query and AAA Reply messages are used for pre-registration procedure to query about MN's profile.

The message between target MAG and LMA ($H_{MAG_LMA}$) includes: $M_{PBU}$ and $M_{PBA}$.

Therefore, the signaling cost of standard handover can be expressed as

$$S^1_{SH} = \frac{P_f}{1-P_f}H_{MN_MAG}(M_3+M_4+M_5+M_6+M_{15}+M_{16}+M_{RS}+M_{RA}) + H_{MAG_MAG}(m(M_7+M_8)+M_9+M_{10}+M_{13}+M_{14}) + H_{MAG_MIIS}(M_3+M_4)+H_{MAG_LMA}(M_{17}+M_{18}+3(M_{PBU}+M_{PBA})) \quad (15)$$

Similarly, we can get that of fast handover solution as follows.

$$S^1_{FH} = \frac{P_f}{1-P_f}H_{MN_MAG}(M_1+M_5+M_6+M_{11}+M_{12}+M_{RS}+M_{RA}) + H_{MAG_MAG}(m(M_7+M_8)+M_9+M_{10}+M_{13}+M_{14}+M_{HI}+M_{HAck}) + H_{MAG_MIIS}(M_3+M_4)+3H_{MAG_LMA}(M_{PBU}+M_{PBA}) \quad (16)$$

The signaling cost of the proposed solution is shown as follows.

$$S^1_{OH} = \frac{P_f}{1-P_f}H_{MN_MAG}(M_1+M_2+M_5+M_6+M_{11}+M_{12}+M_{UNA}) + H_{MAG_MAG}(m(M_7+M_8)+M_{9e}+M_{10e}+M_{13}+M_{14}) + H_{MAG_MIIS}(M_3+M_4)+2H_{MAG_LMA}(M_{PBU}+M_{PBA}) \quad (17)$$

Therefore, we can get the signaling cost based on Eq. (14) - (17).

Figure 13 depicts the signaling cost versus the distance among MAGs ($H_{MAG_MAG}$). The wireless link failure probability is set to 0.5, the cell radius is set to 100m, the speed is set to [1, 50] m/s, and the distance between adjacent roads is set to 10m. We can find that increasing $H_{MAG_MAG}$ results in higher signaling cost. When the $H_{MAG_MAG}$ is less than 8, the standard handover solution has the highest signaling cost than the standard and fat handover solutions. When $H_{MAG_MAG}$ is more than 8, the fast handover solution has the highest signaling cost as it needs additional signaling delivery between the serving MAG and the target MAG to perform the pre-registration. In each case, the proposed solution has the lowest signaling cost as it carries the related information in regular MIH messages.

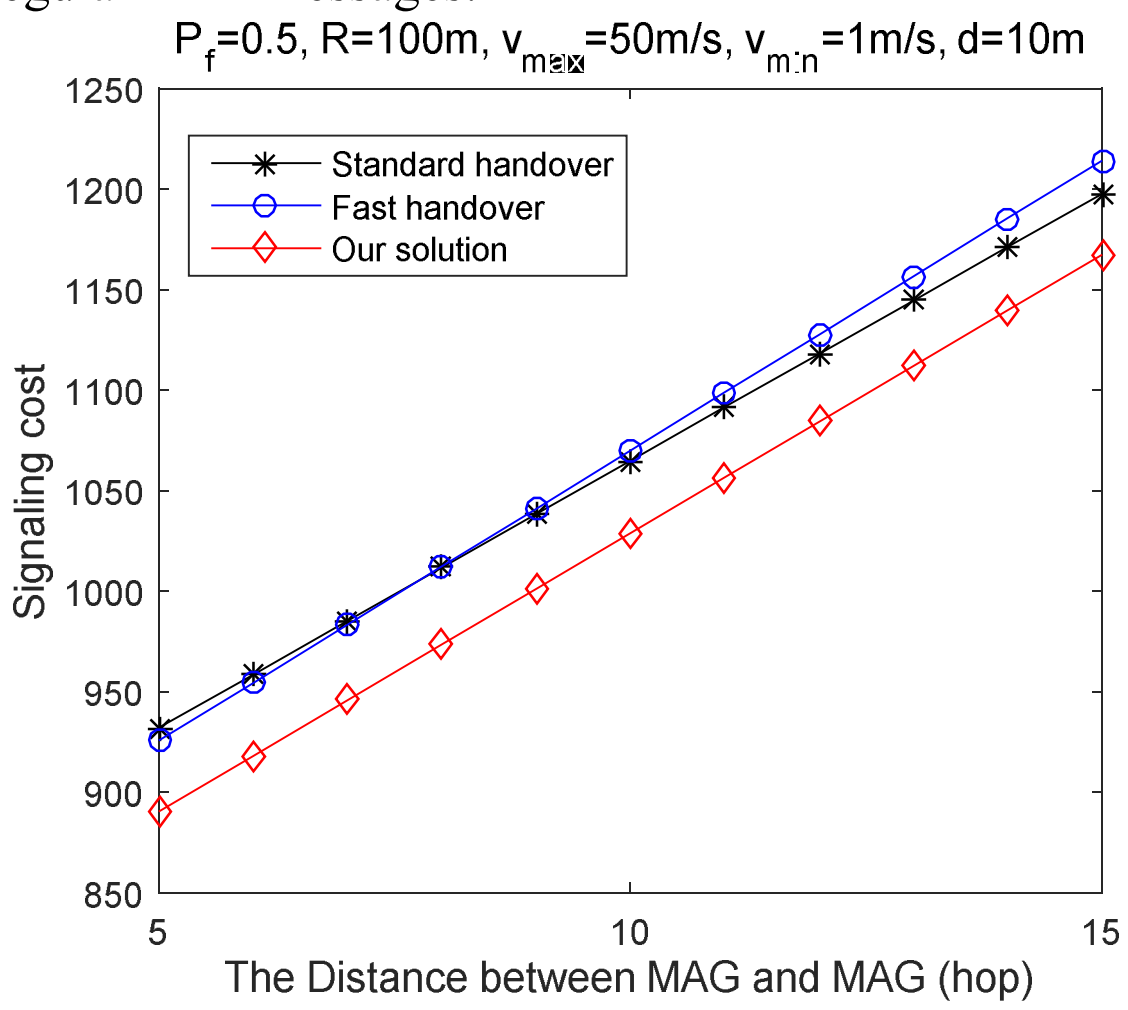

Figure 13 Signaling cost versus the distance among MAGs

Figure 14 illustrates the signaling cost versus wireless link failure probability. The cell radius is set to 100m, the speed is set to [1, 50] m/s, and the distance between adjacent roads is set to 10m. The increasing wireless link failure probability results in higher mobility signaling cost, and the proposed solution has lower signaling cost than the standard handover and fast handover solutions.

Figure 16 shows the signaling cost versus cell radius. The wireless link failure probability is set to 0.5, the cell radius is set to 100m, the speed is set to [1, 50] m/s, and the distance between adjacent roads is set to 10m. The increase of cell radius reduces the probability of handover as well as the signaling cost. The signaling cost of fast handover is slightly higher than the standard handover, while the proposed solution is lower than others.

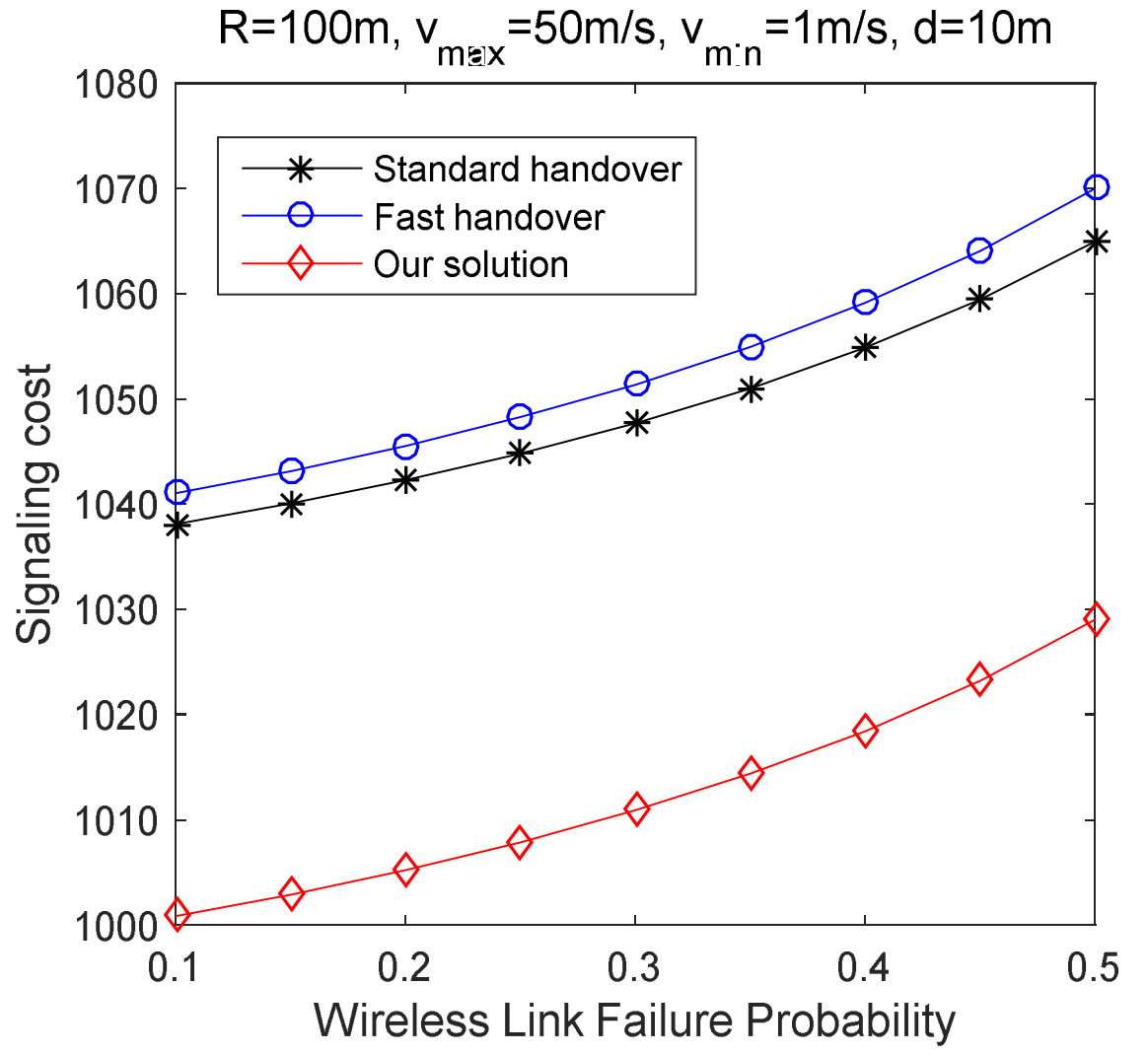

Figure 14 Signaling cost versus wireless link failure probability

Figure 15 shows the signaling cost versus moving speed of MNs. The cell radius is set to 100m and the distance between adjacent roads is set to 10m. The maximum speed is set to 50m/s, and the minimum speed is set to [1, 36]m/s. When MN is moving at high speed, it may cross more cell roads, and incurs more handovers. Accordingly, the signaling cost is increased.

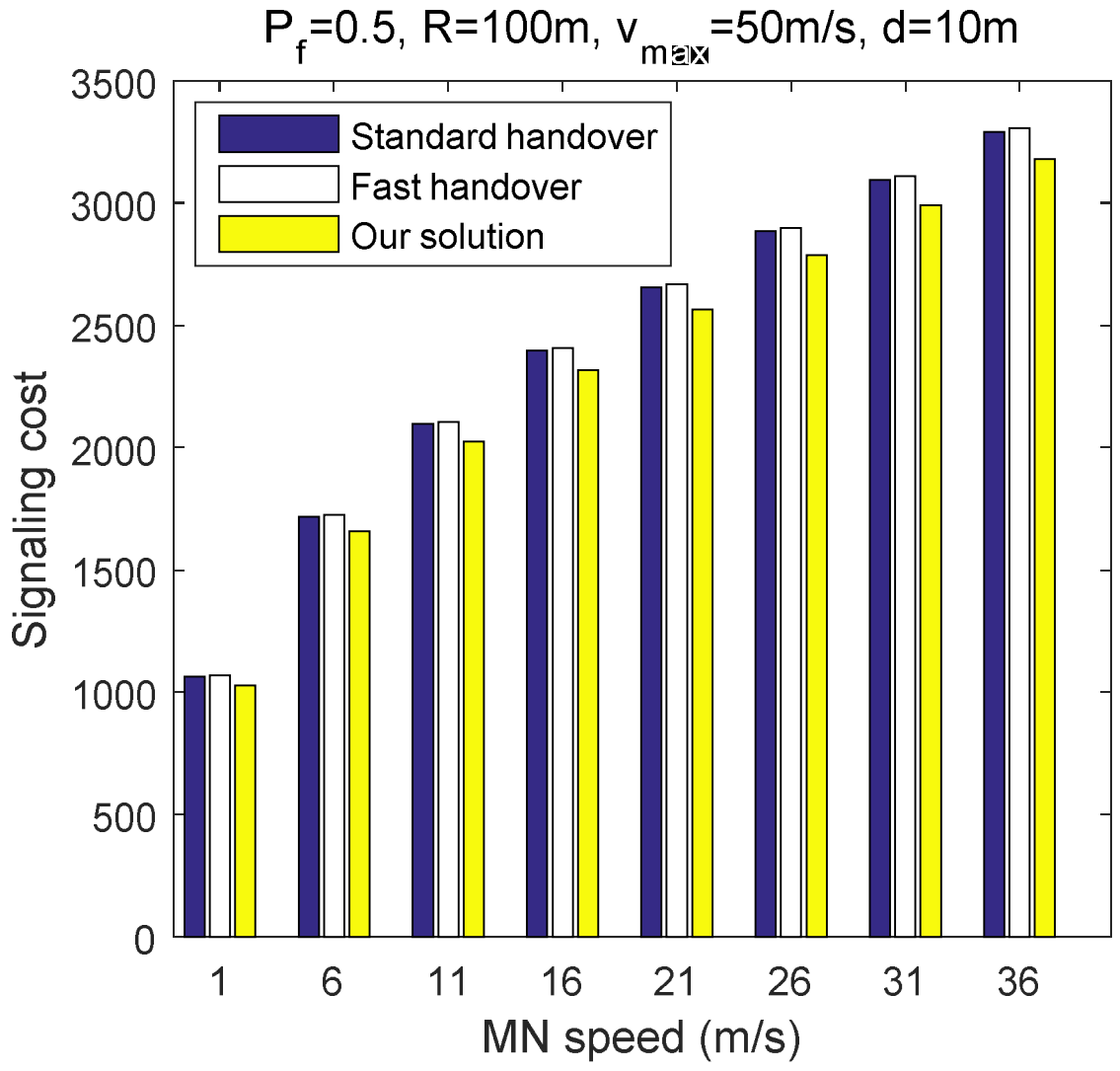

Figure 15 Signaling cost versus moving speed

Similar, as that shown in Figure 16, the signaling cost of the proposed solution is lower than others, and that of fast handover is slightly higher than standard handover.

Figure 17 illustrates the signaling cost versus the distance between adjacent roads. The wireless link failure probability is set to 0.5, the cell radius is set to 100m, the speed is set to [1, 50] m/s. To simplify the analysis, we use $d$ to represent $d_x$ and $d_y$, which can be used as a metric to evaluate the road network density of the city. The increasing of d results in higher signaling cost. This is because the probability of crossing cell is increased when the road network density is sparse as an MN can only move along the road. The signaling cost of the proposed solution is lower than other two, while for fast handover solution, signaling cost is highest.

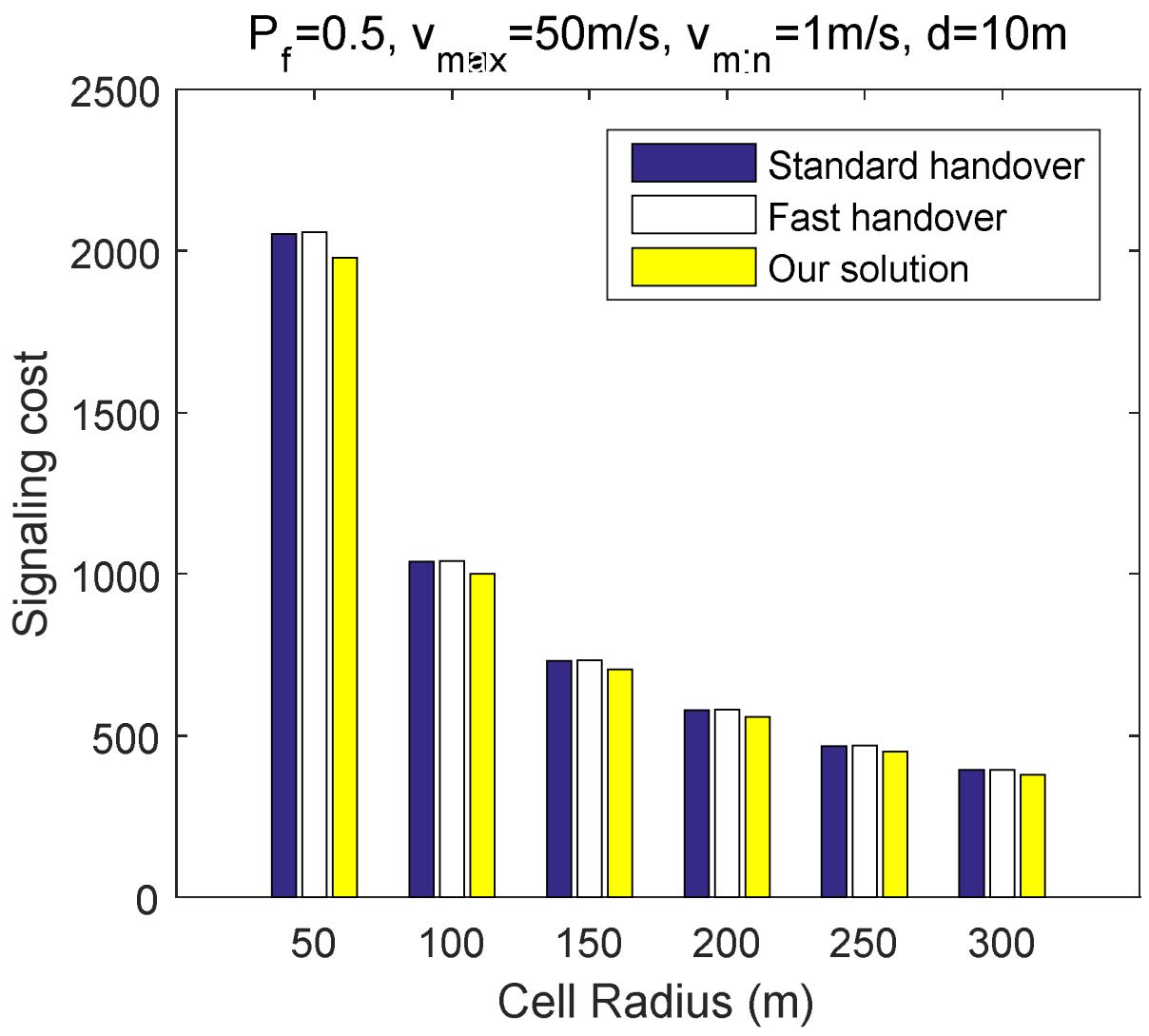

Figure 16 Signaling cost versus cell radius

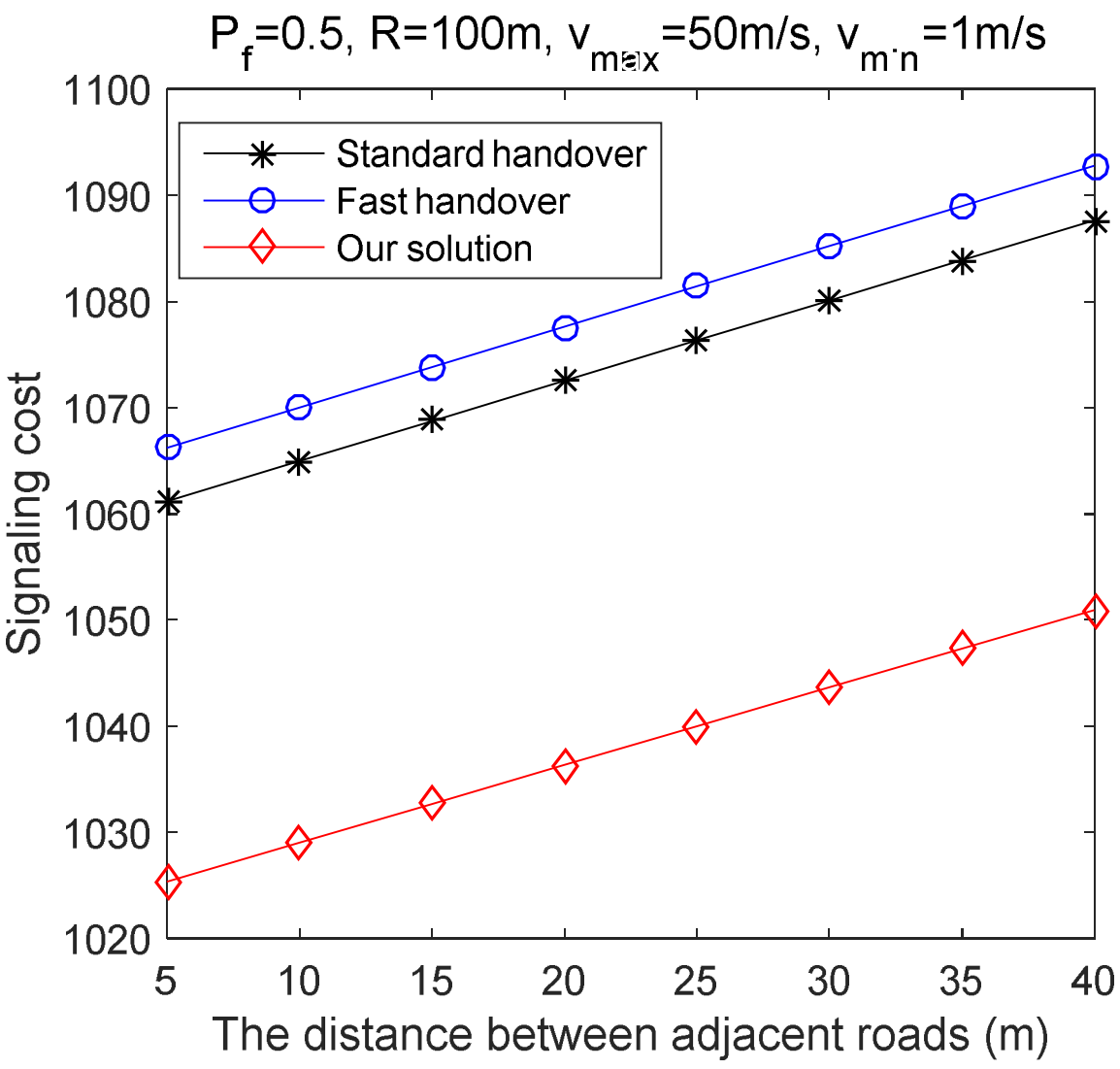

Figure 17 Signaling cost versus the distance between adjacent roads

# 5 Conclusions

This paper proposes an integration model of MIH and FPMIPv6 with the fast handover support, which can reduce the handover delay without introducing additional signaling cost compared with the standard handover solution and the fast handover solution. This is obtained by extending related MIH messages to include the related information for pre-registration, and adopting the transient binding mechanism.

Although the proposed solution can improve the performance of heterogeneous handover, it still has some limitations. It is a Centralized Mobility Management (CMM) which is dependent on the central mobility entity such as HA, Local Mobility Anchor (LMA). However, with the rapid increase of MNs, these central entities may become the single point of failure and result in serious service degradation. To solve this problem, Distributed Mobility Management (DMM) is already available in the literature with the objective to distribute the traffic in more flat architecture with optimal routing. DMM distributes the traffic on different mobility anchors to solve the signal point failure problem, and it can be used in host-based and network-based mobility management. In host-based DMM, it maintains multiple CoAs for each session and sets up the tunnels between MN and its each session's MAs. While in network-based DMM, it maintains multiple tunnels but introducing a CMD to maintain for the sessions for each MN.

This paper uses city section mobility model to evaluate its performance, which is more realistic for its deployment scenarios. The numerical results show that the proposed solution is better than the standard handover solution and the fast handover solution. In the future, we plan to combine the DMM with MIH to analyze their integration model and performance.

# Acknowledgement

This research was partially supported by Basic Science Research Program through the National Research Foundation of Korea (NRF) funded by the Ministry of Science, ICT & Future Planning (2014R1A1A1005915). Also, it was in part supported by the Soonchunhyang University Research Fund.

# References

[1] ICT Facts and Figures, http://www.itu.int/en/ITU-D/Statistics/Pages/facts/default.aspx, 2015.

[2] http://www.cisco.com/c/en/us/solutions/collateral/service-provider/visual-networking-index-vni/mobile-white-paper-c11-520862.pdf

[3] Jianfeng Guan, Changqiao Xu, Hongke Zhang, Huachun Zhou, Mobility Challenges and Management in the Future Wireless Heterogeneous Networks, in Book "Wireless multi-access environments and quality of service provisioning: solutions and application", IGI Global Press, DOI: 10.4018/978-1-4666-0017-1.ch002, 2012.

[4] Raffaele Bolla, Matteo Repetto, A Comprehensive Tutorial for Mobility Management in Data Networks, IEEE Communications Surveys & Tutorials, Vol. 16, No. 2, pp. 812-833, 2014.

[5] C. Perkins, D. Johnson, and J. Arkko, "Mobility Support in IPv6," IETF RFC 6275, July 2011.

[6] H. Soliman, Flarion *et al.*, "Hierarchical Mobile IPv6 Mobility Management (HMIPv6)", RFC 4140, 2005.

[7] Koodli, R., "Mobile IPv6 Fast Handovers", RFC 5568, July 2009, http://tools.ietf.org/pdf/rfc5568.pdf

[8] M. Liehbsch, A. Muhanna, O. Blume, Transient Binding for Proxy Mobile IPv6, RFC 6058, March 2011.

[9] Kim, I., & Kim, Y. (2011). Performance evaluation and improvement of TCP throughput over PFMIPv6 with MIH. In The 6th IFIP/IEEE International Workshop on Broadband Convergence Networks, (pp. 997–1004).

[10] L. K. Indumathi, D. Shalini Punitha Vathani, Shery Radley, An extension of Proxy Mobile IPv6 for Reducing handover delay Latency and Packet Loss using Transient Binding, International Journal of Computer Applications, Vol. 69, No. 15, May 2013.

[11] Vivek Gupta, Subir Das, Anthony Chan, *et al.* IEEE Standard for Local and metropolitan area networks—Media Independent Handover Services, IEEE Std 802.21TM-2008, 2008.

[12] T. Melia, G. Bajko, S. Das, N. Golmie, JC. Zuniga, EEE 802.21 Mobility Services Framework Design (MSFD), RFC 5677, December 2009.

[13] S. Gundavelli et al., "Proxy Mobile IPv6," IETF RFC 5213, Aug. 2008.

[14] Jianfeng Guan, HuaChun Zhou, Weisi Xiao, Zhiwei Yan, Yajuan Qin, Hongke Zhang, Implementation and Analysis of Network-based Mobility Management Protocol in WLAN Environments, In Proceeding of MobiWorld'08, 2008.

[15] Jianfeng Guan, Huachun Zhou, Zhiwei Yan, Yajuan Qin, Hongke Zhang, Implementation and analysis of proxy MIPv6, Wireless Communications and Mobile Computing, Vol. 11, 2011, pp.477-490, 2011

[16] Yokota, H., Chowdhury, K., Koodli, R., Patil, B., & Xia, F. (2010). Fast handovers for proxy mobile ipv6. RFC 5949. http://tools.ietf.org/html/rfc5949.

[17] Murtadha Muayad Khalil, Nor Kamariah Noordin, and Borhanuddin Mohd Ali. "Survey and Analysis of Integrating PMIPv6 and MIH Mobility Management Approaches for Heterogeneous Wireless Networks." Wireless Personal Communications (2015): 1-26.

[18] IEEE Std 802.21 d-2015, IEEE Standard for Local and metropolitan area networks – Part 21: Media Independent Handover Services, Amendment 4: Multicast Group Management, IEEE Standard, 2015.

[19] Kim I., Jung Y., Kim Y. Low latency proactive handover scheme for proxy MIPv6 with MIH. Lecture notes in computer science, pp. 344–353. Berlin: Springer, 2008.

[20] Byun-Kil Kim, Young-Chul Jun, Igor Kim, Young-Tak Kim, Enhance FMIPv4 Horizontal Handover with Minimized Channel Scanning Time based on Media Independent Handover (MIH), In the proceeding of IEEE Network Operations and Management Symposium Workshops, 2008, pp: 52-55.

[21] Kim I., Jung Y., Kim Y. MIH-assisted PFMIPv6 predictive handover with selective channel scanning, fast re-association and efficient tunneling management. Lecture Notes in Computer Science, 5787, 321–330, 2009.

[22] I. Kim and Y. T. Kim, "Prediction-based smart channel scanning with minimized service disruption for IEEE 802.lle WLAN," in IEEE International Conference on Consumer Electronics (ICCE), Las Vegas, USA, Jan. 2011, pp. 934-935.

[23] Jong-Hyouk Lee, Jean-Marie Bonnin, HOTA: Handover Optimized Ticket-Based Authentication in Network-based Mobility Management, Information Sciences, 2013, 64-77.

[24] Jong-Kyouk Lee, Jean-Marie Bonnin, Ilsun You, Tai-Myoung Chung, Comparative Handover Performance Analysis of IPv6 Mobility Management Protocols, IEEE Transcations on Industrial Electronics, 60(3):1077-1088, 2013.

[25] Hossain MS, Atiquzzaman M. Cost analysis of mobility protocols. Telecommun Syst 2013;52(4):2271–85.

[26] J. McNair, I. F. Akyildiz, M. D. Bender, An Inter-System Handoff Technique for the IMT-2000 System, IEEE Infocom 2000, pp. 208-216.

[27] Li Jun Zhang, Samuel Pierre, An enhanced fast handover with seamless mobility support for next-generation wireless networks, Journal of Network and Computer Applications, 46(2014: 322-335.

[28] H. Chan, D. Liu, P. Seite, H. Yokota, J. Korhonen, Requirements for Distributed Mobility Management, RFC 7333, 2014.